\documentclass[a4paper,12pt,leqno]{article}
\usepackage{theorem}
\newtheorem{thm}{THEOREM}
\newtheorem{lem}[thm]{LEMMA}

\theoremheaderfont{\scshape}

\newcommand{\ket}[1]{|#1\rangle}

\title{{\Large {\bf SYMMETRY OF DISTRIBUTION FOR THE ONE-DIMENSIONAL HADAMARD WALK}
}}
\author{{\small NORIO KONNO \footnote{To whom correspondence should be addressed. E-mail: norio@mathlab.sci.ynu.ac.jp} } \\
{\scriptsize Department of Applied Mathematics, 
Faculty of Engineering, 
Yokohama National University}\\
{\scriptsize Hodogaya, Yokohama 240-8501, Japan} \\
{\scriptsize norio@mathlab.sci.ynu.ac.jp}\\
{\small TAKAO NAMIKI} \\
{\scriptsize Division of Mathematics, 
Graduate School of Science, 
Hokkaido University} \\
{\scriptsize Kita, Sapporo 060-0810, Japan} \\
{\scriptsize nami@math.sci.hokudai.ac.jp}\\
{\small TAKAHIRO SOSHI}\\
{\scriptsize Department of Applied Mathematics, 
Faculty of Engineering, 
Yokohama National University}\\
{\scriptsize Hodogaya, Yokohama 240-8501, Japan} \\
{\scriptsize  soshi@lam.osu.sci.ynu.ac.jp}
} 

\date{\empty }
\begin{document}
\maketitle

\par\noindent
\begin{small}
\par\noindent
{\bf Abstract}. In this paper we study a one-dimensional quantum random walk with the Hadamard transformation which is often called the Hadamard walk. We construct the Hadamard walk using a transition matrix on probability amplitude and give some results on symmetry of probability distributions for the Hadamard walk. 

\footnote[0]{
{\it Abbr. title:} One-Dimensional Hadamard walk.}

\end{small}

\setcounter{equation}{0}
\section{Introduction}
\newcommand{\U}{\bar{U}}

Recently, quantum versions of classical random walks have been investigated by a number of groups, for examples, see refs. [1-10], [13-22], [24], [26-29]. A more general setting including quantum cellular automata can be found in ref. [23]. It is well known that classical random walks have found practical applications in many fields including physics, chemistry, biology, astronomy, and computer science. The hope is that quantum random walks might lead to applications unavailable classically. In fact, quantum random walks have many suggested applications in physics and quantum information. For examples, quantum walks have the potential to give various new tools for quantum algorithm design [3]. It is also believed that quantum walks can provide a benchmarking protocol for ion trap quantum computers [28]. Childs $et\ al.$ [9] shows query complexity of some problems is speed up exponentially by quantum walk. Kempe [13] and Yamasaki $et\ al.$ [29] are the first which showed exponential speed up of hitting time in discrete time walk, while Childs $et\ al.$ [9] treats continuous time model. (Yamasaki $et\ al.$ [29] conjectured the speed up by numerical computation, and the conjecture was proven by Kempe [13].) So to clarify some properties of the distribution for the quantum random walk can be considered as important work. 
\par

In the present paper, we focus on the one-dimensional Hadamard walk whose time evolution is given by the Hadamard transformation (see p.19 in ref. [25]):
\begin{eqnarray*}
H=\frac{1}{\sqrt2}
\left[
\begin{array}{cc}
1 & 1 \\
1 &-1 
\end{array}
\right]
\end{eqnarray*}
\par\noindent
The Hadamard matrix $H$ is unitary. The Hadamard walk considered here is a quantum generalization of a classical symmetric random walk in one dimension with an additional degree of freedom called the chirality. The chirality takes values left and right, and means the direction of the motion of the particle. The evolution of the Hadamard walk is given by the following way. At each time step, if the particle has the left chirality, it moves one step to the left, and if it has the right chirality, it moves one step to the right.
\par
More precisely, the Hadamard matrix $H$ acts on two chirality states $\ket{L}$ and $\ket{R}$:
\begin{eqnarray*}
&& \ket{L} \>\> \to \>\> {1 \over \sqrt{2}}(\ket{L} + \ket{R}) \\
&& \ket{R} \>\> \to \>\> {1 \over \sqrt{2}}(\ket{L} - \ket{R}) 
\end{eqnarray*}
where $L$ and $R$ refer to the right and left chirality state respectively. In fact, define
\begin{eqnarray*}
\ket{L} = 
\left[
\begin{array}{cc}
1 \\
0  
\end{array}
\right],
\qquad
\ket{R} = 
\left[
\begin{array}{cc}
0 \\
1  
\end{array}
\right]
\end{eqnarray*}
so we have
\begin{eqnarray*}
&& H\ket{L} = {1 \over \sqrt{2}}(\ket{L} + \ket{R}) \\
&& H\ket{R} = {1 \over \sqrt{2}}(\ket{L} - \ket{R}) 
\end{eqnarray*}
Here we introduce $P$ and $Q$ matrices as follows:
\begin{eqnarray*}
P=\frac{1}{\sqrt2}
\left[
\begin{array}{cc}
1 & 1 \\
0 & 0 
\end{array}
\right], 
\quad
Q=\frac{1}{\sqrt2}
\left[
\begin{array}{cc}
0 & 0 \\
1 & -1 
\end{array}
\right]
\end{eqnarray*}
with 
\[
H=P+Q
\]
Here $P$ (resp. $Q$) represents that the particle moves to the left (resp. right) with equal probability. We should remark that $P$ and $Q$ are useful tools in the study of the iterates of $H$. However, they cannot be interpreted as dynamical evolution operators since they are not unitary. By using $P$ and $Q$, we define the dynamics of the Hadamard walk in one dimension. To do so, we introduce the next $(2N+1) \times (2N+1)$ matrix $\overline{H}_N : ({\bf C}^2)^{2N+1} \to ({\bf C}^2)^{2N+1}$ :
\begin{eqnarray*}
\overline{H}_N =
\left[
\begin{array}{ccccccc}
0 & P & 0 & \dots & \dots & 0 & Q \\
Q & 0 & P & 0 & \dots & \dots & 0 \\
0 & Q & 0 & P & 0 &\dots &0\\
\vdots & \ddots & \ddots & \ddots & \ddots & \ddots & \vdots \\
0 & \dots & 0 & Q & 0 & P & 0\\
0 & \dots & \dots & 0 & Q & 0 & P\\
P & 0 & \dots & \dots & 0 & Q & 0
\end{array}
\right]
\end{eqnarray*}
where ${\bf C}$ is the set of complex numbers and 
\[
0 = 
\left[
\begin{array}{cc}
0 & 0 \\
0 & 0 
\end{array}
\right]
\]
Let 
\[
\Psi_k ^{(n)} (\varphi) = 
\left[
\begin{array}{cc}
\Psi_{L,k} ^{(n)} (\varphi) \\
\Psi_{R,k} ^{(n)} (\varphi)   
\end{array}
\right] = 
\Psi_{L,k} ^{(n)} (\varphi) \ket{L} 
+ \Psi_{R,k} ^{(n)} (\varphi) \ket{R} \in {\bf C}^2
\]
be the two component vector of amplitudes of the particle being at site $k$ and at time $n$ with the chirality being left (upper component) and right (lower component), and 
\[
\Psi^{(n)} (\varphi) =  {}^t[\Psi_{-N} ^{(n)} (\varphi), \Psi_{-(N-1)} ^{(n)} (\varphi), \ldots, \Psi_{N} ^{(n)} (\varphi)]
\in ({\bf C}^2)^{2N+1}
\]
be the qubit states at time $n$ where $t$ means transposed operator. Here the initial qubit state is given by 
\[
\Psi^{(0)} (\varphi) = {}^t[\overbrace{0, \ldots , 0}^N, \varphi, \overbrace{0, \ldots, 0}^N ] \in ({\bf C}^2)^{2N+1}
\]
where 
\[
0 =
\left[
\begin{array}{cc}
0 \\
0   
\end{array}
\right], \quad
\varphi =
\left[
\begin{array}{cc}
\alpha \\
\beta   
\end{array}
\right]
\]
with $|\alpha|^2 + |\beta|^2=1$ and $|z|$ is the absolute value of $z \in {\bf C}.$ The following equation defines the time evolution of the Hadamard walk:
\begin{eqnarray*} 
(\overline{H}_N \Psi^{(n)} (\varphi))_k = Q \Psi_{k-1} ^{(n)} (\varphi) + P 
\Psi_{k+1} ^{(n)} (\varphi) 
\end{eqnarray*}
where $(\Psi^{(n)} (\varphi))_k = \Psi_k ^{(n)} (\varphi)$. Note that $P$ and $Q$ satisfy
\begin{eqnarray*}
PP^*+QQ^* = P^* P + Q^*Q=I, \> PQ^*=QP^*=Q^*P=P^*Q=
\left[
\begin{array}{cc}
0 & 0 \\
0 & 0  
\end{array}
\right]
\end{eqnarray*}
where $*$ means the adjoint operator. The above relations imply that $\overline{H}_N$ becomes also unitary matrix. 

\par
For initial state $\overline{\varphi}= {}^t[\overbrace{0,\ldots,0}^N, \varphi , \overbrace{0,\ldots,0}^N]$, we have 
\begin{eqnarray*}
&& \overline{H}_N \overline{\varphi} = {}^t[\overbrace{0,\dots , 0}^{N-1},P \varphi,0,Q \varphi,\overbrace{0,\dots,0}^{N-1}]\\
&& \overline{H}_N^2 \overline{\varphi} = {}^t[\overbrace{0,\ldots,0}^{N-2},P^2\varphi,0,(PQ+QP)\varphi,0,Q^2\varphi,\overbrace{0, \ldots,0}^{N-2}]\\
&& \overline{H}_N^3 \overline{\varphi} = {}^t[\overbrace{0,\ldots,0}^{N-3},P^3\varphi,0,(P^2Q+PQP+QP^2)\varphi,0,(Q^2P+QPQ+PQ^2)\varphi,0,Q^3\varphi,\overbrace{0,\ldots,0}^{N-3}]
\end{eqnarray*}
This shows that expansion of $H^n = (P+Q)^n$ for the Hadamard walk corresponds to that of $2^n = (1 + 1)^n$ for symmetric classical random walk, that is, Pascal's triangle.

\par
Define the set of initial qubit states as follows:
\[
\Phi = \left\{ \varphi =
\left[
\begin{array}{cc}
\alpha \\
\beta   
\end{array}
\right]
\in 
{\bf C}^2
:
|\alpha|^2 + |\beta|^2 =1
\right\}
\]
Now we define probability distribution of the Hadamard walk $X_n ^{\varphi}$ starting from initial qubit state $\varphi \in \Phi$ by 
\begin{eqnarray*}
P(X_n ^{\varphi} = k) = |\Psi_k ^{(n)} (\varphi)|^2
\end{eqnarray*}

The purpose of this paper is to study some properties of distribution of $X_n ^{\varphi}$, for example, the symmetry of distribution. To do so, we introduce the following classes of initial qubit states:
\begin{eqnarray*}
\Phi_{\bot} &=& \left\{ \varphi = 
\left[
\begin{array}{cc}
\alpha \\
\beta 
\end{array}
\right] 
\in 
\Phi :
|\alpha|= |\beta|, \> \alpha \overline{\beta} + \overline{\alpha} \beta =0 
\right\} \\
\Phi_s &=&  \{ \varphi \in 
\Phi : \> 
P(X_n ^{\varphi}=k) = P(X_n ^{\varphi}=-k) \>\>
\hbox{for any} \> n \in {\bf Z}_+ \> \hbox{and} \> k \in {\bf Z}
\} \\
\Phi_0 &=& \left\{ \varphi \in 
\Phi : \> 
E(X_n ^{\varphi})=0 \>\> \hbox{for any} \> n \in {\bf Z}_+
\right\}
\end{eqnarray*}
and ${\bf Z}$ (resp. ${\bf Z}_+$) is the set of (resp. non-negative) integers.  We should remark that if $\alpha \beta \not= 0$, then $\alpha \overline{\beta} + \overline{\alpha} \beta =0$ implies that there is a $k \in {\bf Z}$ such that $\theta_{\alpha}=\theta_{\beta}+(\pi/2)+k\pi$, where $\overline{z}$ is conjugate of $z \in {\bf C}$ and $\theta_{z}$ is the argument of $z \in {\bf C}$, that is, $\alpha$ and $\beta$ are orthogonal. For $\varphi \in \Phi_s$, the distribution of $X_n ^{\varphi}$ is symmetric for any $n \in {\bf Z}_+$. 

In this setting, we have the next result (Theorem 2 in Section 2):
\[
\Phi_{\bot} = \Phi_s = \Phi_0
\]
This result can be generalized to a general quantum random walk in one dimension (see Konno [17, 18], for example). However here we give other two proofs which appear in Sections 2 and 3 respectively.

\par
The rest of the paper is organized as follows. In Section 2, we study the necessary and sufficient conditions of the symmetry of distribution of $X_n ^{\varphi}$ for any time step $n$. Section 3 treats another approach to the study of the symmetry by using a quantum version of Pascal's triangle (see Theorem 3). Section 4 is devoted to a conjecture on the expectation of the Hadamard walk. Appendix A gives a proof of Theorem 3. In Appendix B, we compute the $m$th moment of a limit distribution for the rescaled Haramard walk in the symmetric case.

\section{Symmetry of Distribution}
First we present the following useful lemma to prove Theorem 2. Let
\[
J=
\left[
\begin{array}{cc}
0 & -1 \\
1 & 0 
\end{array}
\right]
\]
Then we have
\begin{lem}
\label{lem:lem1}
\par\noindent
\rmfamily{(i)} We suppose that initial qubit state is 
\[
\varphi 
= \frac{e^{i \theta}}{\sqrt2}
\left[
\begin{array}{cc}
1 \\
i 
\end{array}
\right]
\]
where $\theta \in [0, 2 \pi)$. Then we have
\begin{eqnarray}
\Psi_k^{(n)} (\varphi)  = (-1)^{n} i J \Psi_{-k}^{(n)} (\varphi)
\end{eqnarray}
for any $k \in {\bf Z}$ and $n \in {\bf Z}_{+}$.
\par\noindent
\hbox{(ii)} We suppose that initial qubit state is 
\[
\varphi 
= \frac{e^{i \theta}}{\sqrt2}
\left[
\begin{array}{cc}
1 \\
-i 
\end{array}
\right]
\]
where $\theta \in [0, 2 \pi)$. Then we have
\begin{eqnarray}
\Psi_k^{(n)} (\varphi) = (-1)^{n} (-i) J \Psi_{-k}^{(n)} (\varphi)
\end{eqnarray}
for any $k \in {\bf Z}$ and $n \in {\bf Z}_{+}$.
\end{lem}
\par
\
\par\noindent
Proof. We show parts (i) and (ii) by induction on time step $n$. The proof of part (ii) is essentially the same as that of part (i), so we omit it. For simplicity, we write $\Psi_k^{(n)} = \Psi_k^{(n)} (\varphi).$
\par
First we consider $n=0$ case. For any $k \not= 0$, the initial condition gives $\Psi_k^{(0)}= 0$, so we can check Eq. (2.1). As for $k=0$, we have
\[
(-1)^{0} i J \Psi_{0}^{(0)} = i 
\left[
\begin{array}{cc}
0 & -1 \\
1 & 0 
\end{array}
\right] 
\frac{e^{i \theta}}{\sqrt2}
\left[
\begin{array}{cc}
1 \\
i  
\end{array}
\right]
= \Psi_{0}^{(0)} 
\]
Therefore Eq. (2.1) is correct for $n=0.$
\par 
Next we assume that Eq. (2.1) holds for time $n=m$. We should remark that the definition of the Hadamard walk gives 
\begin{eqnarray}
\Psi_k^{(m+1)}=Q\Psi_{k-1}^{(m)}+P\Psi_{k+1}^{(m)}
\end{eqnarray}
for any $k$. Moreover note that 
\begin{eqnarray}
QJ=-JP, \quad PJ=-JQ
\end{eqnarray}
Then we have
\begin{eqnarray*}
\Psi_{k}^{(m+1)} 
&=& Q \Psi_{k-1}^{(m)}+P\Psi_{k+1}^{(m)} \\
&=& Q (-1)^m iJ \Psi_{-(k-1)}^{(m)}+P (-1)^m iJ \Psi_{-(k+1)}^{(m)} \\
&=& (-1)^m i \{ QJ \Psi_{-(k-1)}^{(m)}+ PJ \Psi_{-(k+1)}^{(m)} \} \\
&=& (-1)^{m+1} i J \{ P \Psi_{-(k-1)}^{(m)}+ Q \Psi_{-(k+1)}^{(m)} \} \\
&=& (-1)^{m+1} iJ \Psi_{-k}^{(m+1)}
\end{eqnarray*}
The first and fifth equalities are given by Eq. (2.3). The second equality comes from the induction hypothesis. The fourth equality is obtained by Eq. (2.4). So it is shown that Eq. (2.1) is also correct for time $n=m+1$. The proof of Lemma 1 is complete.
\par
\
\par
The following three classes were given in Introduction.
\begin{eqnarray*}
\Phi_{\bot} &=& \left\{ \varphi = 
\left[
\begin{array}{cc}
\alpha \\
\beta 
\end{array}
\right] 
\in 
\Phi :
|\alpha|= |\beta|, \> \alpha \overline{\beta} + \overline{\alpha} \beta =0 
\right\} \\
\Phi_s &=&  \left\{ \varphi \in 
\Phi : \> 
| \Psi_k^{(n)} (\varphi)| = | \Psi_{-k} ^{(n)} (\varphi)|
\> \hbox{for any} \> n \in {\bf Z}_+ \> \hbox{and} \> k \in {\bf Z}
\right\} \\
\Phi_0 &=& \left\{ \varphi \in 
\Phi : \> 
E(X_n ^{\varphi})=0 \> \hbox{for any} \> n \in {\bf Z}_+
\right\}
\end{eqnarray*}
As we mentioned in Introduction, it is noted that $\alpha \overline{\beta} + \overline{\alpha} \beta =0$ with $\alpha \beta \not= 0$ implies that $\alpha$ and $\beta$ are orthogonal. For $\varphi \in \Phi_s$, the probability distribution of $X_n ^{\varphi}$ is symmetric for any $n \in {\bf Z}_+$. By using Lemma 1, we obtain the following result:
\begin{thm}
\label{thm:thm2} For the Hadamard walk, we have
\begin{eqnarray*}
\Phi_{\bot} = \Phi_s = \Phi_0
\end{eqnarray*}
\end{thm}
\par
\
\par\noindent
Proof. First we see that the definitions of $\Phi_s$ and $\Phi_0$ give
\begin{eqnarray}
\Phi_s \subset \Phi_0 
\end{eqnarray}
\par
Next we should remark that 
\[
\Phi_{\bot} = \left\{
\frac{e^{i \theta_1}}{\sqrt2}
\left[
\begin{array}{cc}
1 \\
i 
\end{array}
\right],
\> \> 
\frac{e^{i \theta_2}}{\sqrt2}
\left[
\begin{array}{cc}
1 \\
-i 
\end{array}
\right]
\> : \theta_1, \theta_2 \in [0, 2 \pi) \right\}.
\]
By the above remark and Lemma 1, for any $\varphi \in \Phi_{\bot}$, we have
\[
| \Psi_k^{(n)} (\varphi)|^2 =(-1)^{2n} |i|^2 \Psi_{-k} ^{(n)} (\varphi)^{\ast} 
J^{\ast} J \Psi_{-k} ^{(n)} (\varphi) = | \Psi_{-k} ^{(n)} (\varphi)|^2
\]
Therefore if $\varphi \in \Phi_{\bot}$, then 
\[
| \Psi_k^{(n)} (\varphi)| = | \Psi_{-k} ^{(n)} (\varphi)|
\]
for any $n \in {\bf Z}$ and $k \in {\bf Z}_+$, so we have
\begin{eqnarray}
\Phi_{\bot} \subset \Phi_s
\end{eqnarray}
A direct computation gives
\begin{eqnarray*}
E(X_1) &=& E(X_2) = -  (\alpha \overline{\beta} + \overline{\alpha} \beta ) 
\\
E(X_3) &=& {1 \over 2}(|\beta|^2 - |\alpha|^2) - 
( \alpha \overline{\beta} + \overline{\alpha} \beta )
\end{eqnarray*}
The above equations imply that if $\varphi \in \Phi_0$ then $\varphi \in \Phi_{\bot}$. So we have 
\begin{eqnarray}
\Phi_0 \subset \Phi_{\bot}
\end{eqnarray}
Combining Eqs. (2.5)-(2.7) gives
\[
\Phi_{\bot} \subset \Phi_s \subset \Phi_0 \subset \Phi_{\bot}
\]
so the proof of Theorem 2 is complete.
\par
\
\par\noindent
\section{Quantum Pascal's Triangle}
In this section, we take another approach to the study on symmetry of distribution of $X_n ^{\varphi}$, more precisely, distribution of $P(X_n ^{\varphi} =k)$ for $n+k=$ even. To do so, we need to consider the following expression. For fixed $l$ and $m$ with $l+m=n$ and $m-l=k$, 
\begin{eqnarray*}
\Xi(l,m)= \sum_{l_j, m_j \ge 0: m_1+ \cdots +m_n=m, l_1+ \cdots +l_n=l} P^{l_1}Q^{m_1}P^{l_2}Q^{m_2} \cdots P^{l_n}Q^{m_n}
\end{eqnarray*}
For example, in the case of $P(X_4 = -2)$, we need to know the expression:
\[
\Xi (3,1) = P^3Q+P^2QP+PQP^2+QP^3
\]
We remark the next useful relation:
\[
P^2 = {1 \over \sqrt{2}}P, \qquad Q^2 = -{1 \over \sqrt{2}}Q
\]
By using these, $\Xi (3,1)$ becomes
\[
\Xi (3,1) = {1 \over 2} PQ + {1 \over \sqrt{2}}PQP+{1 \over \sqrt{2}}PQP+{1 \over 2} QP
\]
Here we introduce other useful matrices:
\[
R=\sqrt{2} PQ=\frac{1}{\sqrt2}
\left[
\begin{array}{cc}
1 & -1 \\
0 & 0 
\end{array}
\right], 
\quad
S=\sqrt{2}QP=
\frac{1}{\sqrt2}
\left[
\begin{array}{cc}
0 & 0 \\
1 & 1 
\end{array}
\right].
\]
In general, we obtain the next table of computations with $P,Q,R$ and $S$:
\par
\
\par
\begin{center}
\begin{tabular}{c|cccc}
  & $P$ & $Q$ & $R$ & $S$  \\ \hline
$P$ & $P$ & $R$ & $R$ & $P$  \\
$Q$ & $S$ & $-Q$& $Q$ & $-S$ \\
$R$ & $P$ & $-R$& $R$ & $-P$ \\
$S$ & $S$ & $Q$ & $Q$ & $S$ 
\end{tabular}
\end{center}
where we omit $1/\sqrt2$ factor, for example, $PQ=R/\sqrt2$. Since $P, Q, R$ and $S$ are a basis for the set of $2 \times 2$ matrices with complex valued components, we have
\begin{eqnarray*}
\Xi (l,m) = p_n (l,m) P + q_n (l,m) Q + r_n (l,m) R + s_n (l,m) S
\end{eqnarray*}
Next problem is to obtain explicit forms of $p_n (l,m), q_n (l,m), r_n (l,m)$ and $s_n (l,m)$. In the above example of $n=l+m=4$ case, we have
\begin{eqnarray*}
&& \Xi (4,0) = \left( {1 \over \sqrt2} \right)^3 P, \quad
\Xi (3,1) = \left( {1 \over \sqrt2} \right)^3 (2P+R+S), \quad
\Xi (2,2) = \left( {1 \over \sqrt2} \right)^3 (-P+Q), \\
&& \Xi (1,3) = \left( {1 \over \sqrt2} \right)^3 (-2Q+R+S), \quad
\Xi (2,2) = -\left( {1 \over \sqrt2}\right) ^3 Q
\end{eqnarray*}
For a general $\Xi (l,m)$, the following result is obtained. 
\par
\
\par
\begin{thm}
\label{thm:thm3} We consider the Hadamard walk. Suppose that $l,m \ge 0$ with $l+m=n$, then we have 
\par\noindent
\hbox{(i)} for $l \wedge m (= \min \{ l, m \}) \ge 1$,  
\begin{eqnarray*}
\Xi (l,m) = \left( {1 \over \sqrt2} \right)^{n-1} (-1)^m 
\sum_{\gamma =1} ^{l \wedge m} (-1)^{\gamma} {l-1 \choose \gamma- 1} 
{m-1 \choose \gamma- 1} 
\left[ {l- \gamma \over \gamma } P - {m- \gamma \over \gamma} Q + R + S \right]
\end{eqnarray*}
\par\noindent
\hbox{(ii)} for $l (=n) \ge 1, m = 0$,  
\begin{eqnarray*}
\Xi (l,0) = \left( {1 \over \sqrt2} \right)^{l-1} P
\end{eqnarray*}
\par\noindent
\hbox{(iii)} for $l = 0, m (=n) \ge 1$,  
\begin{eqnarray*}
\Xi (0,m) = \left({1 \over \sqrt2} \right)^{m-1} (-1)^{m+1}Q
\end{eqnarray*}
\end{thm}
Theorem 3 can be obtained as a special case of Lemma 1 in Konno [18]. However for the convenience of readers, we give the proof in Appendix A.
\par
\
\par
From now on we will give an application of this theorem as follows. By Theorem 3, we can get $\Phi_{\bot} \subset \Phi_s$. For $l \wedge m \ge 1$, 
\begin{eqnarray*}
&& {}^t\Xi (l,m) \Xi (l,m) \\
&& = \left( {1 \over 2} \right)^{n-1}  
\sum_{\gamma =1} ^{l \wedge m} \sum_{\delta =1} ^{l \wedge m} (-1)^{\gamma + \delta} {l-1 \choose \gamma- 1} {l-1 \choose \delta- 1} {m-1 \choose \gamma- 1} {m-1 \choose \delta - 1} \\
&&
\qquad \qquad \times ^t\left[ {l- \delta \over \delta } P - {m- \delta \over \delta} Q + R + S \right]
\left[ {l- \gamma \over \gamma } P - {m- \gamma \over \gamma} Q + R + S \right] \\
&& = 
\left( {1 \over 2} \right)^{n}  
\sum_{\gamma =1} ^{l \wedge m} \sum_{\delta =1} ^{l \wedge m} {(-1)^{\gamma + \delta} \over \gamma \delta} {l-1 \choose \gamma- 1} {l-1 \choose \delta- 1} {m-1 \choose \gamma- 1} {m-1 \choose \delta - 1} \\
&& 
\qquad \qquad \times \left[
\begin{array}{cc}
l^2 + (m-2 \gamma)(m -2 \delta) & l(l-2 \gamma)-m(m-2 \delta) \\
l(l-2 \delta)-m(m-2 \gamma) & m^2 + (l-2 \gamma)(l -2 \delta) 
\end{array}
\right].
\end{eqnarray*}
To consider the symmetry, for $l \wedge m \ge 1$, we have to check
\begin{eqnarray*}
&& {}^t\Xi (l,m) \Xi (l,m) - {}^t\Xi (m,l) \Xi (m,l) \\
&& =  
\left( {1 \over 2} \right)^{n}  
\sum_{\gamma =1} ^{l \wedge m} \sum_{\delta =1} ^{l \wedge m} {(-1)^{\gamma + \delta} \over \gamma \delta} {l-1 \choose \gamma- 1} {l-1 \choose \delta- 1} {m-1 \choose \gamma- 1} {m-1 \choose \delta - 1}  \\
&&
\qquad \qquad \times 2(l-m) \left[
\begin{array}{cc}
\gamma + \delta & n - 2 (\gamma + \delta) \\
n - 2 (\gamma + \delta) & - ( \gamma + \delta) 
\end{array}
\right]
\end{eqnarray*}
Similarly, for $l \ge 1$ and $m=0$, we have
\[ 
{}^t\Xi (l,0) \Xi (l,0) - {}^t\Xi (0,l) \Xi (0,l)
= \left( {1 \over 2} \right)^{n-1} \left(^tP P - ^tQ Q \right)
=  \left( {1 \over 2} \right)^{n-1} 
\left[
\begin{array}{cc}
0 & 1 \\
1 & 0 
\end{array}
\right] 
\]
In any case, for any $l, m \ge 0$ except $l=m=0$, there exist $a, b \in {\bf C}$ such that 
\[
{}^t\Xi (l,m) \Xi (l,m) - {}^t\Xi (m,l) \Xi (m,l) =
\left[
\begin{array}{cc}
a & b \\
b & -a 
\end{array}
\right] 
\]
Therefore we obtain
\[
| \Psi_k ^{(n)} (\varphi)|^2 - | \Psi_{-k} ^{(n)} (\varphi)|^2 = 
[ \overline{\alpha}, \overline{\beta}] 
\left[
\begin{array}{cc}
a & b \\
b & -a 
\end{array}
\right] 
\left[
\begin{array}{cc}
\alpha  \\
\beta
\end{array}
\right] 
= a (|\alpha|^2 - |\beta|^2) +b ( \alpha \overline{\beta} + \overline{\alpha} \beta)
\]
where $k=m-l$. From the above result, we see that if $\varphi ={}^t[\alpha, \beta] \in \Phi_{\bot}$, then $\varphi \in \Phi_s$ immediately.

\section{Discussions}
In the case of the Hadamard walk, a direct computation implies that for $n \in \{1,2, \ldots ,10 \}$, 
\begin{eqnarray*}
E(X_n ^{\varphi}) = - a_n (|\alpha|^2 - |\beta|^2 ) - b_n (\alpha \overline{\beta} + \overline{\alpha} \beta)
\end{eqnarray*}
where
\begin{eqnarray*}
&& a_1=a_2=0, \> a_3={1 \over 2}, \> a_4=1, \> a_5={9 \over 8}, \> a_6={5 \over 4}, \> a_7 = {27 \over 16}, \> a_8={17 \over 8}, \> a_9={293 \over 128}, \> 
a_{10}={157 \over 64} \\
&& b_1=b_2=b_3=1, \> b_4={3 \over 2}, \> b_5=2, \> b_6={17 \over 8}, \> b_7 = {9 \over 4}, \> b_8={43 \over 16}, \> b_9 ={25 \over 8}, \> b_{10}={421 \over 128}
\end{eqnarray*}
So we conjecture that $b_{n+1} = a_n +1$ for any $n \ge 1$. This property would be useful for studying symmetry of distribution.
\par
Recently Konno [17, 18] gave a new type of limit theorems for quantum random walks in one dimension. In the symmetric case, his result implies that $X_n ^{\varphi}/n$ converges weakly to $Z^{\varphi}$ whose $m$th moment (= $c_m$) is given by
\begin{eqnarray*}
E((Z^{\varphi})^{2n}) = 1 - {1 \over \sqrt{2}} \sum_{k=0} ^{n-1} 
{1 \over 2^{3k}} { 2k \choose k }, \qquad 
E((Z^{\varphi})^{2n-1}) = 0
\end{eqnarray*}
for any $n \ge 1$ (a derivation of the $m$th moment can be seen in Appendix B). By using finite moments $c_m$ ($m=0,1, \ldots$), we obtain the associated orthogonal polynomials $\{ P_m \}$. So the interacting  Fock space corresponding to the limit distribution for our symmetric Hadamard walk can be derived from the usual way (for example, see pp.73-76 in Hashimoto, Hora and Obata [12]). To clarify the relation between the quantum random walk and the quantum decomposition is one of the future interesting problems.
\par
\
\par\noindent
{\bf Appendix A: Proof of Theorem 3}
\par
Here we give a proof of Theorem 3.
\par
\
\par\noindent
(a) $p_n (l,m)$ case : First we assume $l \ge 2$ and $m \ge 1$. In order to compute $p_n (l,m)$, it is enough to consider only the following case:
\begin{eqnarray*}
C(p,w)^{(2 \gamma +1)} _n (l,m) = \overbrace{PP \cdots P}^{w_1} \overbrace{QQ \cdots Q}^{w_2} \overbrace{PP \cdots P}^{w_3} \cdots \overbrace{QQ \cdots Q}^{w_{2 \gamma}} \overbrace{PP \cdots P}^{w_{2 \gamma+1}} 
\end{eqnarray*}
where $w=(w_1, w_2, \ldots , w_{2 \gamma+1}) \in {\bf Z}_+ ^{2 \gamma +1}$ with  $w_1, w_2, \ldots , w_{2 \gamma+1} \ge 1$ and $\gamma \ge 1$. For example, $PQP$ case is $w_1=w_2=w_3=1$ and $\gamma =1$. We should remark that $l$ is the number of $P$'s and $m$ is the number of $Q$'s, so we have
\begin{eqnarray*}
&& l= w_1 + w_3 + \cdots + w_{2 \gamma +1} \\
&& m= w_2 + w_4 + \cdots + w_{2 \gamma}
\end{eqnarray*}
Moreover $2 \gamma +1$ means the number of clusters of $P$'s and $Q$'s. Next we consider the range of $\gamma$. The minimum is $\gamma =1$, that is, 3 clusters. This case is
\[
P \cdots P Q \cdots Q P \cdots P 
\]
The maximum is $\gamma = (l-1) \wedge m$. This case is
\[
PQPQPQ \cdots PQPQ P \cdots P \>\> (l-1 \ge m), \quad PQPQPQ \cdots PQ P Q \cdots Q P \>\> (l-1 \le m)
\]
for examples. Here we introduce a set of sequences with $2 \gamma +1$ components: for fixed $\gamma \in [1, (l-1) \wedge m]$ 
\begin{eqnarray*}
W(p,2 \gamma +1) 
&=& \{ w = (w_1, w_2, \cdots , w_{2 \gamma +1}) \in {\bf Z}_+ ^{2 \gamma +1} : 
w_1 + w_3 + \cdots + w_{2 \gamma +1} =l, \\ 
&& \qquad \qquad  w_2 + w_4 + \cdots + w_{2 \gamma} =m, \> w_1, w_2, \ldots, w_{2\gamma}, w_{2 \gamma +1} \ge 1 \} 
\end{eqnarray*}
From the next relations:
\begin{eqnarray*}
P^2 = {1 \over \sqrt{2}} P, \quad Q^2 = -{1 \over \sqrt{2}}Q, \quad PQ= {1 \over \sqrt{2}} R, \quad R^2= {1 \over \sqrt{2}} R, \quad RP={1 \over \sqrt{2}} P
\end{eqnarray*}
we have 
\begin{eqnarray*}
C(p,w)_n ^{(2 \gamma +1)}(l,m) 
&=& \left({1 \over \sqrt{2}} \right)^{w_1-1} P \left(-{1 \over \sqrt{2}} \right)^{w_2 -1}Q \cdots \left(-{1 \over \sqrt{2}} \right)^{w_{2 \gamma} -1} Q \left({1 \over \sqrt{2}} \right)^{w_{2 \gamma +1} -1} P \\
&=& \left({1 \over \sqrt{2}} \right)^{l-(\gamma +1)} \left(-{1 \over \sqrt{2}} \right)^{m - \gamma} (PQ)^{\gamma} P \\
&=& \left({1 \over \sqrt{2}}\right)^{l-(\gamma +1)} \left(-{1 \over \sqrt{2}}\right)^{m - \gamma} \left({1 \over \sqrt{2}}\right)^{\gamma} R^{\gamma} P \\ 
&=& \left({1 \over \sqrt{2}} \right)^{l-(\gamma +1)} \left(-{1 \over \sqrt{2}} \right)^{m - \gamma} \left({1 \over \sqrt{2}} \right)^{\gamma} \left({1 \over \sqrt{2}} \right)^{\gamma-1} RP \\ 
&=& \left({1 \over \sqrt{2}} \right)^{n - 1} (-1)^{m + \gamma} P
\end{eqnarray*}
where $w \in W(p, 2 \gamma +1)$. For $l \ge 2,$ and $m \ge 1$, that is, $\gamma \ge 1$, we obtain
\begin{eqnarray*}
C(p,w)_n ^{(2 \gamma +1)}(l,m) = 
\left({1 \over \sqrt{2}} \right)^{n - 1} (-1)^{m + \gamma} P
\end{eqnarray*}
Note that the right hand side of the above equation does not depend on $w \in W(p, 2 \gamma +1)$. Finally we compute the number of $w=(w_1,w_2, \ldots, w_{2\gamma}, w_{2 \gamma +1})$ satisfying $w \in  W(p, 2 \gamma +1)$ by a standard combinatorial argument as follows:
\[
|W(p, 2 \gamma +1)| = { l-1 \choose \gamma } { m-1 \choose \gamma -1}
\]
From the above observation, we obtain
\begin{eqnarray*}
p_n (l,m) P
&=& \sum_{\gamma =1} ^{(l-1) \wedge m}
\sum_{w \in W(p,2 \gamma +1)}
C(p,w)_n ^{(2 \gamma +1)}(l,m) \\
&=&
\sum_{\gamma =1} ^{(l-1) \wedge m}
|W(p,2 \gamma +1)| C(p,w)_n ^{(2 \gamma +1)}(l,m) \\
&=&
\sum_{\gamma =1} ^{(l-1) \wedge m}
{ l-1 \choose \gamma } { m-1 \choose \gamma -1} 
\left({1 \over \sqrt{2}} \right)^{n - 1} (-1)^{m + \gamma} P
\end{eqnarray*}
Therefore the desired result is obtained:
\begin{eqnarray*}
p_n (l,m) = \left({1 \over \sqrt{2}} \right)^{n - 1} (-1)^m
\sum_{\gamma =1} ^{(l-1) \wedge m} (-1)^{\gamma} { l-1 \choose \gamma } { m-1 \choose \gamma -1} 
\end{eqnarray*}
When $l \ge 1$ and $m=0$, this case is
\begin{eqnarray*}
p_n (l,0) P = P^l = \left({1 \over \sqrt{2}} \right)^{l-1} P
\end{eqnarray*}
so we have
\begin{eqnarray*}
p_n (l,0)= \left({1 \over \sqrt{2}} \right)^{l-1} 
\end{eqnarray*}
Furthermore, when $l=1, m \ge 1$ and $l=0, m \ge 0$, we see  
\begin{eqnarray*}
p_n (l,m) = 0
\end{eqnarray*}
(b) $q_n (l,m)$ case : As in the case of part (a), we assume $l \ge 1$ and $m \ge 2$. In order to compute $q_n (l,m)$, it is sufficient to consider only the following case:
\begin{eqnarray*}
C(q,w)^{(2 \gamma +1)} _n (l,m) = \overbrace{QQ \cdots Q}^{w_1} \overbrace{PP \cdots P}^{w_2} \overbrace{QQ \cdots Q}^{w_3} \cdots \overbrace{PP \cdots P}^{w_{2 \gamma}} \overbrace{QQ \cdots Q}^{w_{2 \gamma+1}} 
\end{eqnarray*}
where $w=(w_1, w_2, \ldots , w_{2 \gamma+1}) \in {\bf Z}_+ ^{2 \gamma +1}$ with $w_1, w_2, \ldots , w_{2 \gamma+1} \ge 1$ and $\gamma \ge 1$. For example, $QPQ$ case is $w_1=w_2=w_3=1$ and $\gamma =1$. Then we have
\begin{eqnarray*}
&& l= w_2 + w_4 + \cdots + w_{2 \gamma} \\
&& m= w_1 + w_3 + \cdots + w_{2 \gamma +1}
\end{eqnarray*}
As in the previous part, the range of $\gamma$ is between 1 and $l \wedge (m-1)$. Here we introduce a set of sequences with $2 \gamma +1$ components: for fixed $\gamma \in [1, l \wedge (m-1)]$ 
\begin{eqnarray*}
W(q,2 \gamma +1) 
&=& \{ w = (w_1, w_2, \cdots , w_{2 \gamma +1}) \in {\bf Z}_+ ^{2 \gamma +1} : 
w_2 + w_4 + \cdots + w_{2 \gamma} =l, \\ 
&& \qquad \qquad  w_1 + w_3 + \cdots + w_{2 \gamma +1} =m, \> w_1, w_2, \ldots, w_{2\gamma}, w_{2 \gamma +1} \ge 1 \} 
\end{eqnarray*}
Similarly we have 
\begin{eqnarray*}
C(q,w)_n ^{(2 \gamma +1)}(l,m) 
= \left({1 \over \sqrt{2}} \right)^{n - 1} (-1)^{m + \gamma + 1} Q
\end{eqnarray*}
for $l \ge 1, m \ge 2$ and $w \in W(q, 2 \gamma +1)$. Furthermore
\[
|W(q, 2 \gamma +1)| = { l-1 \choose \gamma -1} { m-1 \choose \gamma }
\]
From these, we obtain
\begin{eqnarray*}
q_n (l,m) Q
&=& \sum_{\gamma =1} ^{(l-1) \wedge m}
\sum_{w \in W(q,2 \gamma +1)}
C(q,w)_n ^{(2 \gamma +1)}(l,m) \\
&=&
\sum_{\gamma =1} ^{l \wedge (m-1)}
|W(q,2 \gamma +1)| C(q,w)_n ^{(2 \gamma +1)}(l,m) \\
&=&
\left({1 \over \sqrt{2}} \right)^{n - 1} (-1)^{m+1} 
\sum_{\gamma =1} ^{l \wedge (m-1)} (-1)^{\gamma} 
{ l-1 \choose \gamma -1} { m-1 \choose \gamma} Q
\end{eqnarray*}
Therefore the desired result is obtained:
\begin{eqnarray*}
q_n (l,m) =
- \left({1 \over \sqrt{2}} \right)^{n - 1} (-1)^{m} 
\sum_{\gamma =1} ^{l \wedge (m-1)} (-1)^{\gamma} 
{ l-1 \choose \gamma -1} { m-1 \choose \gamma} 
\end{eqnarray*}
When $l = 0$ and $m \ge 1$, this case is
\begin{eqnarray*}
q_n (0,m)= \left(-{1 \over \sqrt{2}} \right)^{m-1} 
\end{eqnarray*}
Furthermore, when $l \ge 1, m = 1$ and $l \ge 0, m = 0$, we see  
\begin{eqnarray*}
q_n (l,m) = 0
\end{eqnarray*}
(c) $r_n (l,m)$ case : First we assume $l \ge 1$ and $m \ge 1$. In order to compute $r_n (l,m)$, we consider only the following case:
\begin{eqnarray*}
C(r,w)^{(2 \gamma)} _n (l,m) = \overbrace{PP \cdots P}^{w_1} \overbrace{QQ \cdots Q}^{w_2} \overbrace{PP \cdots P}^{w_3} \cdots \overbrace{QQ \cdots Q}^{w_{2 \gamma}}
\end{eqnarray*}
where $w=(w_1, w_2, \ldots , w_{2 \gamma}) \in {\bf Z}_+ ^{2 \gamma}$ with $w_1, w_2, \ldots , w_{2 \gamma } \ge 1$ and $\gamma \ge 1$. For example, $PQPQ$ case is $w_1=w_2=w_3=w_4=1$ and $\gamma =2$. Note that
\begin{eqnarray*}
&& l= w_1 + w_3 + \cdots + w_{2 \gamma -1} \\
&& m= w_2 + w_4 + \cdots + w_{2 \gamma}
\end{eqnarray*}
Moreover $2 \gamma$ means the number of clusters of $P$'s and $Q$'s. Next we consider the range of $\gamma$. The minimum is $\gamma =1$, that is, 2 clusters. This case is
\[
P \cdots P Q \cdots Q 
\]
The maximum is $\gamma = l \wedge m$. This case is
\[
PQPQPQ \cdots PQP \cdots PQ \>\> (l \ge m), \quad PQPQPQ \cdots PQ PQ \cdots Q  \>\> (m \ge l)
\]
for examples. Here we introduce a set of sequences with $2 \gamma$ components: for fixed $\gamma \in [1, l \wedge m]$ 
\begin{eqnarray*}
W(r,2 \gamma) 
&=& \{ w = (w_1, w_2, \cdots , w_{2 \gamma}) \in {\bf Z} ^{2 \gamma} : 
w_1 + w_3 + \cdots + w_{2 \gamma -1} =l, \\ 
&& \qquad \qquad  w_2 + w_4 + \cdots + w_{2 \gamma} =m, \> w_1, w_2, \ldots, w_{2\gamma -1}, w_{2 \gamma} \ge 1 \} 
\end{eqnarray*}
As in the case of part (a), Eq. (3.9) becomes 
\begin{eqnarray*}
C(r,w)_n ^{(2 \gamma)}(l,m) 
= \left({1 \over \sqrt{2}} \right)^{n - 1} (-1)^{m + \gamma } R
\end{eqnarray*}
for $l \ge 1, m \ge 1$, that is, $\gamma \ge 1$, and $w \in W(r, 2 \gamma)$. Furthermore
\[
|W(r, 2 \gamma)| = { l-1 \choose \gamma -1} { m-1 \choose \gamma -1}
\]
From the above results, we obtain
\begin{eqnarray*}
r_n (l,m) R
&=& \sum_{\gamma =1} ^{l \wedge m}
\sum_{w \in W(r,2 \gamma)}
C(r,w)_n ^{(2 \gamma)}(l,m) \\
&=&
\sum_{\gamma =1} ^{l \wedge m}
|W(r,2 \gamma )| C(r,w)_n ^{(2 \gamma)}(l,m) \\
&=&
\left({1 \over \sqrt{2}} \right)^{n - 1} (-1)^m
\sum_{\gamma =1} ^{l \wedge m}
(-1)^{\gamma}
{ l-1 \choose \gamma -1} { m-1 \choose \gamma -1} R
\end{eqnarray*}
Therefore 
\begin{eqnarray*}
r_n (l,m) =
\left({1 \over \sqrt{2}} \right)^{n - 1} (-1)^m
\sum_{\gamma =1} ^{l \wedge m}
(-1)^{\gamma}
{ l-1 \choose \gamma -1} { m-1 \choose \gamma -1}
\end{eqnarray*}
When $l \wedge m =0$, this case is
\begin{eqnarray*}
r_n (l,m) = 0 
\end{eqnarray*}
(d) $s_n (l,m)$ case : First we assume $l \ge 1$ and $m \ge 1$. In order to compute $s_n (l,m)$, it is enough to consider only the following case:
\begin{eqnarray*}
C(s,w)^{(2 \gamma)} _n (l,m) = \overbrace{QQ \cdots Q}^{w_1} \overbrace{PP \cdots P}^{w_2} \overbrace{QQ \cdots Q}^{w_3} \cdots \overbrace{PP \cdots P}^{w_{2 \gamma}}
\end{eqnarray*}
where $w=(w_1, w_2, \ldots , w_{2 \gamma}) \in {\bf Z}_+ ^{2 \gamma}$ with $w_1, w_2, \ldots , w_{2 \gamma } \ge 1$ and $\gamma \ge 1$. For example, $QPQP$ case is $w_1=w_2=w_3=w_4=1$ and $\gamma =2$. Then we have
\begin{eqnarray*}
&& l= w_2 + w_4 + \cdots + w_{2 \gamma} \\
&& m= w_1 + w_3 + \cdots + w_{2 \gamma -1}
\end{eqnarray*}
As in the case of part (c), we obtain
\begin{eqnarray*}
s_n (l,m) =
\left({1 \over \sqrt{2}} \right)^{n - 1} (-1)^m
\sum_{\gamma =1} ^{l \wedge m}
(-1)^{\gamma}
{ l-1 \choose \gamma -1} { m-1 \choose \gamma -1}
\end{eqnarray*}
When $l \wedge m =0$, this case is
\begin{eqnarray*}
s_n (l,m) = 0 
\end{eqnarray*}
It should be noted that 
\[
r_n (l,m) = s_n (l,m)
\]
for any $l$ and $m$. When $l \wedge m \ge 1$, combining the above results (a) - (d) gives the proof of part (i). Proofs of parts (ii) and (iii) are trivial. So the proof of Theorem 3 is complete.
\par
\
\par\noindent
{\bf Appendix B: Computation of the $m$th Moment}
\par\noindent
We begin with
\begin{eqnarray*}
E((Z^{\varphi})^{2n})
=
\int^{1/\sqrt{2}} _{-1/\sqrt{2}} 
{x^{2n} \over \pi (1-x^2) \sqrt{1-2x^2}} dx
=
{2^{(3-2n)/2} \over \pi} \times I_{2n}
\end{eqnarray*}
where $Z^{\varphi}$ is a weak limit distribution of the rescaled Hadamard walk starting from $\varphi$ and
\begin{eqnarray*}
I_n 
= \int^1 _0 {y^{n} \over (2-y^2) \sqrt{1-y^2}} dy 
= \int^{\pi/2} _0 {\sin ^{n} \theta \over 1+ \cos^2 \theta} d \theta 
\end{eqnarray*}
It should be noted that 
\begin{eqnarray*}
I_{2n+2}
&=& 
\int^{\pi/2} _0 { \{ 2 - (1+\cos^2 \theta) \} \sin ^{2n} \theta
\over 1+ \cos^2 \theta} d \theta \\ 
&=&
2 I_{2n} - {(2n-1)(2n-3) \cdots 1 \over 2n(2n-2) \cdots 2} 
\times {\pi \over 2}
\end{eqnarray*}
Let $J_{n} = I_{2n} / 2^n$. Then we see that for $n \ge 0$
\begin{eqnarray*}
J_{n+1} - J_n
= - {\pi \over 2^{3n+2}} {2n \choose n}
\end{eqnarray*}
where
\begin{eqnarray*}
J_0 ={\pi \over 2 \sqrt{2}}, \qquad J_1 = {2 - \sqrt{2} \over 4 \sqrt{2}} \pi
\end{eqnarray*}
So we have
\begin{eqnarray*}
I_{2n} = 2^n J_n 
= 2^n \left[ {1 \over 2 \sqrt{2}} - \sum_{k=0} ^{n-1} 
{1 \over 2^{3k+2}} {2k \choose k} \right] \pi
\end{eqnarray*}
Therefore for $n \ge 1$, 
\begin{eqnarray*}
E((Z^{\varphi})^{2n})
=
1 - {1 \over \sqrt{2}} \sum_{k=0} ^{n-1} 
{1 \over 2^{3k}} {2k \choose k} 
\end{eqnarray*}
In particular,
\begin{eqnarray*}
E((Z^{\varphi})^{2})
=
{2 - \sqrt{2} \over 2}
\end{eqnarray*}

\par
\
\par\noindent
{\bf Acknowledgment.}  This work is partially financed by the Grant-in-Aid for Scientific Research (B) (No.12440024) of Japan Society of the Promotion of Science.
\par
\
\par
\begin{small}
\bibliographystyle{jplain}

\end{small}

\end{document}